\pdfoutput=1
\documentclass[twocolumn,pra, aps,superscriptaddress,showpacs]{revtex4-1}
\usepackage[utf8]{inputenc}
\usepackage[colorlinks=true,citecolor=myblue,linkcolor=myblue,urlcolor=myblue]{hyperref}
\usepackage{mathptmx}
\usepackage{subfigure}
\usepackage{dcolumn}
\usepackage{amsmath,amssymb}
\usepackage{bm}
\usepackage{xspace}
\usepackage{color}
\usepackage{latexsym}
\usepackage{color}
\usepackage[english]{babel}
\usepackage{psfrag,graphicx}
\usepackage{graphicx}
\usepackage{epsf}
\usepackage{amsmath}
\usepackage{amssymb}
\usepackage{amsfonts}
\usepackage{bm}
\usepackage{natbib}
\usepackage{epstopdf}
\usepackage{epstopdf}
\usepackage{appendix}
\usepackage[export]{adjustbox}

\definecolor{mygrey}{gray}{0.35}
\definecolor{myblue}{rgb}{0.2,0.2,0.8}
\definecolor{myzard}{cmyk}{0,0,0.05,0}
\definecolor{mywhite}{rgb}{1,1,1}
\definecolor{myred}{rgb}{1,0.,0.3}

\DeclareMathAlphabet{\mathcal}{OMS}{cmsy}{m}{n}

\def\fm#1{\ifmmode #1 \else $#1$\fi}
\def\ket#1{{%
  \ifmmode |\,#1\,\rangle \else $|\,#1\,\rangle$\fi}}
\def\bra#1{{%
  \ifmmode \langle\,#1\,| \else $\langle\,#1\,|$\fi}}
\def\braket#1#2{{%
  \ifmmode \langle\,#1\,|\,#2\,\rangle \else $\langle\,#1\,|\,#2\,\rangle$\fi}}

\def\Al{\fm{\mathrm{Al}^{+}}\xspace}
\def\Alts{\fm{^{27}\Al}\xspace}
\def\Lu{\fm{\mathrm{Lu}^{+}}\xspace}
\def\Luoss{\fm{^{176}\mathrm{Lu}^{+}}\xspace}

\def\Ba{\fm{\mathrm{Ba}^{+}}\xspace}
\def\Baote{\fm{^{138}\Ba}\xspace}

\def\Sr{\fm{\mathrm{Sr}^{+}}\xspace}
\def\Sree{\fm{^{88}\Sr}\xspace}
\def\Ca{\fm{\mathrm{Ca}^{+}}\xspace}
\def\Caf{\fm{^{40}\Ca}\xspace}
\def\Yb{\fm{\mathrm{Yb}^{+}}\xspace}

\def\In{\fm{\mathrm{In}^{+}}\xspace}

\def\dsoh{\fm{{}^2\mathrm{S}_{1/2}}\xspace}

\def\ddfh{\fm{{}^2\mathrm{D}_{5/2}}\xspace}

\def\dsohddfh{\dsoh\fm{\leftrightarrow}\ddfh\xspace}

\def\be{\begin{equation}}
\def\ee{\end{equation}}
\def\ba{\begin{align}}
\def\enda{\end{align}}
\def\bi{\begin{itemize}}
\def\ei{\end{itemize}}

 \def\ee{\mathord{\rm e}}

 \def\ee{\mathord{\rm e}}

\renewcommand{\ee}{{\rm e}}

\def\beq{\begin{equation}}
\def\beq{\begin{equation}}
\def\eeq{\end{equation}}

\begin{document}

\title[Short Title]{Robust optical clock transitions in trapped ions}

\author{Nati Aharon}
\affiliation{Racah Institute of Physics, The Hebrew University of Jerusalem, Jerusalem
91904, Givat Ram, Israel}
\author{Nicolas Spethmann}
\affiliation{QUEST Institute for Experimental Quantum Metrology, Physikalisch-Technische Bundesanstalt, 38116 Braunschweig, Germany}
\author{Ian D. Leroux}
\affiliation{QUEST Institute for Experimental Quantum Metrology, Physikalisch-Technische Bundesanstalt, 38116 Braunschweig, Germany}
\affiliation{National Research Council Canada, Ottawa, Ontario, K1A 0R6, Canada}
\author{Piet O. Schmidt}
\affiliation{QUEST Institute for Experimental Quantum Metrology, Physikalisch-Technische Bundesanstalt, 38116 Braunschweig, Germany}
\affiliation{Institut für Quantenoptik, Leibniz Universität Hannover, 30167 Hannover, Germany}
\author{Alex Retzker}
\affiliation{Racah Institute of Physics, The Hebrew University of Jerusalem, Jerusalem
91904, Givat Ram, Israel}

\date{\today}

\begin{abstract}
We present a novel method for engineering an optical clock transition that is robust against external field fluctuations and is able to overcome limits resulting from field inhomogeneities. The technique is based on the application of continuous driving fields to form a pair of dressed states essentially free of all relevant shifts. Specifically, the clock transition is robust to magnetic shifts, quadrupole and other tensor shifts, and amplitude fluctuations of the driving fields. The scheme is applicable to either a single ion or an ensemble of ions, and is relevant for several types of ions, such as \Caf, \Sree, \Baote and \Luoss. Taking a spherically symmetric Coulomb crystal formed by 400 \Caf ions as an example, we show through numerical simulations that the inhomogeneous linewidth of tens of Hertz in such a crystal together with linear Zeeman shifts of order 10~MHz are reduced to form a linewidth of around 1~Hz. We estimate a two-order-of-magnitude reduction in averaging time compared to state-of-the art single ion frequency references, assuming a probe laser fractional instability of $10^{-15}$. Furthermore, a statistical uncertainty reaching $2.9\times 10^{-16}$ in 1~s is estimated for a cascaded clock scheme in which the dynamically decoupled Coulomb crystal clock stabilizes the interrogation laser for an \Alts clock.
\end{abstract}
%\pacs{03.67.Ac, 03.67.Pp}
\maketitle

\section{Introduction}
Optical clocks based on neutral atoms trapped in optical lattices and single trapped ions have reached estimated systematic uncertainties of a few parts in $10^{-18}$ \cite{ludlow_optical_2015, nicholson_systematic_2015, huntemann_single-ion_2016, poli_optical_2013}. Taking advantage of these record uncertainties for applications ranging from relativistic geodesy \cite{muller_high_2018, grotti_geodesy_2018, mehlstaubler_atomic_2018, takano_geopotential_2016} over fundamental physics \cite{safronova_search_2018, kozlov_highly_2018, delva_test_2017} to frequency metrology \cite{denker_geodetic_2017, lisdat_clock_2016, grebing_realization_2016, nemitz_frequency_2016, yamanaka_frequency_2015} requires achieving statistical measurement uncertainties of the same level within practical averaging times $\tau$ (given in seconds). This has been achieved with single-ensemble optical lattice clocks in self-comparison experiments up to a level of $1.6\times 10^{-16}/\sqrt{\tau}$ \cite{al-masoudi_noise_2015} and by implementing an effectively dead-time-free clock consisting of two independent clocks probed in an interleaved fashion \cite{schioppo_ultrastable_2017}, reaching a statistical uncertainty of $0.6\times 10^{-16}/\sqrt{\tau}$. 
In contrast to neutral atom lattice clocks, which are typically probed with hundreds to thousands of atoms, single ion clocks are currently limited in their statistical uncertainty by quantum projection noise \cite{itano_quantum_1993} to levels of a few parts in $10^{-15}/\sqrt{\tau}$ \cite{huntemann_single-ion_2016, chou_frequency_2010, dube_$88mathrmsr+$_2015}. The statistical uncertainty can be improved by probing for longer times, ultimately limited by the excited clock state lifetime or the laser coherence time \cite{peik_laser_2006, leroux_-line_2017}. Alternatively, the number of probed ions can be increased.

Recently, multi-ion clock schemes have been proposed to address this issue \cite{keller_optical_2017,arnold_prospects_2015,herschbach_linear_2012,champenois_ion_2010}. However, several challenges have to be overcome to maintain and transfer the small and very well characterizable systematic shifts achievable with single trapped ions to larger ion crystals. The oscillating rf field in Paul traps results in ac Stark and second order Doppler shifts through micromotion \cite{berkeland_minimization_1998, keller_precise_2015, itano_external-field_2000}. Furthermore, electric field gradients from the trapping fields and the surrounding ions couple to atomic quadrupole moments, resulting in an electric quadrupole shift (QPS). The effects of micromotion can be avoided by trapping strings of ions in a precision machined linear Paul trap with negligible excess micromotion from trap imperfections \cite{herschbach_linear_2012, pyka_high-precision_2014}. The QPS in such chains can be avoided by choosing an ion species with negligible differential electric quadrupole moment between the clock states, such as \In or \Al, or by employing ring traps in which the QPS is the same for all ions \cite{champenois_ion_2010}. A high-accuracy multi-ion clock based on ion chains containing on the order of tens of ions in a linear quadrupole trap has been proposed and is expected to achieve trap-induced fractional systematic uncertainties at the $10^{-19}$ level \cite{keller_optical_2017}.

An alternative approach based on large 3d Coulomb crystals of ions has been theoretically investigated \cite{arnold_prospects_2015} for ion species for which the micromotion-induced Doppler shift and the ac Stark shift, both driven by the rf trapping field, can be made to cancel at a ``magic'' rf drive frequency \cite{berkeland_minimization_1998}. This cancellation has been employed for single \Sr \cite{dube_high-accuracy_2014}, single \Ca \cite{huang_frequency_2016}, and is currently being investigated for \Lu \cite{arnold_prospects_2015, paez_atomic_2016, arnold_suppression_2016}. However, electronic states with $J>1/2$ are subject to rank 2 tensor shifts, such as rf electric field-induced tensor ac Stark (TASS) and QPS \cite{berkeland_minimization_1998,itano_external-field_2000}. For clouds of 100s of ions this results in position-dependent shifts, since the electric field environment of the ions differ. It has been proposed to reduce this inhomogeneous broadening across the ion crystal by employing a spherical ion crystal to minimize the QPS and by adding a compensating laser field for the TASS \cite{arnold_prospects_2015}, or by operating at a judiciously chosen magnetic-field-insensitive point \cite{arnold_suppression_2016} for ions with hyperfine structure.

\begin{figure}[t]
\centering{}\includegraphics[width=1\linewidth,right]{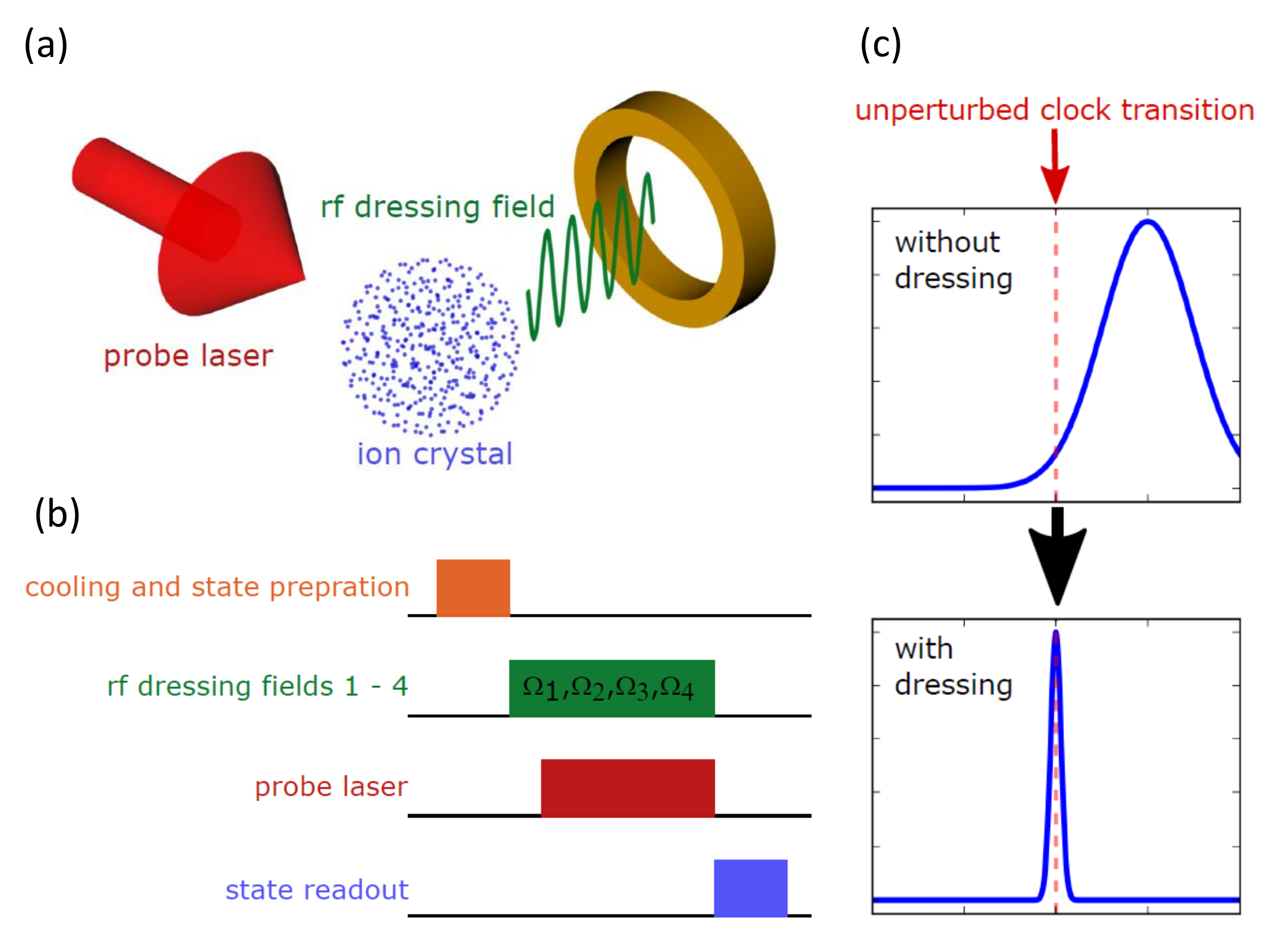}\protect\caption{Schematic representation of the setup. (a) The ion crystal is probed while it is being manipulated by the rf dressing fields. (b) Schematic representation of the sequence considered in this scheme. (c) Illustration of the robust clock transition. While the unperturbed clock transition is subject to magnetic shifts, quadrupole shifts, and tensor shifts, the dressing fields substantially mitigate these shifts and result in a robust clock transition.}\label{Setup}
\end{figure}

Achieving insensitivity of atomic energy levels to external field fluctuations has been theoretically investigated using pulsed or continuous dynamical decoupling (CDD) \cite{viola_dynamical_1998, gordon_optimal_2008, cai_robust_2012, cohen_continuous_2017, fanchini_continuously_2007, aharon_fully_2016, cai_robust_2012, bermudez_robust_2012, zanon-willette_magic_2012, kazakov_magic_2015,aharon_general_2013,aharon_general_2017}. Such schemes have been experimentally implemented in various systems ranging from NV centres and solid state spin systems \cite{laucht_dressed_2017, farfurnik_experimental_2017, golter_protecting_2014, barfuss_strong_2015, rohr_synchronizing_2014, souza_experimental_2012, lange_universal_2010,stark_high_freq_2017,stark_clock_2018} to neutral atoms \cite{jones_rydberg_2013} and trapped ions \cite{biercuk_optimized_2009, baumgart_ultrasensitive_2016, noguchi_generation_2012-1, timoney_quantum_2011, webster_simple_2013, manovitz_fast_2017}.

% non-resonant CDD protocols related to clocks
CDD or dressed-state engineering in the context of clocks has been proposed for linear Zeeman shift cancellation in neutral atom clocks through rf dressing of Zeeman substates in a regime where the drive Rabi frequency is larger than the drive frequency \cite{zanon-willette_magic_2012}. However, this scheme is unable to cancel tensorial shifts, such as the QPS or tensor Stark shifts. In a similar approach, magnetic field noise suppression up to second order in radio-frequency clocks via weak rf dressing has been proposed \cite{kazakov_magic_2015} and experimentally implemented to engineer synthetic clock states for rf spectroscopy in ultra-cold rubidium \cite{sarkany_controlling_2014, trypogeorgos_synthetic_2018}. 

Here, we show through numerical simulations that CDD using four rf frequencies significantly suppresses all relevant homogeneous (linear Zeeman) and inhomogeneous (micromotion-induced second order Doppler, QPS, scalar and tensor ac Stark) frequency shifts on an optical clock transition for ion crystals containing on the order of 100 ions or more. We demonstrate the basic principle, which is illustrated in Fig. \ref{Setup}, on the \dsohddfh clock transition in \Caf, but the scheme is directly applicable to other systems as well.  It therefore allows the operation of a multi-ion clock \cite{keller_optical_2017} using ion species whose clock transitions have a non-vanishing differential electric quadrupole moment.  One of the many possible applications of such a multi-ion frequency reference is the phase stabilization of a probe laser for a single ion clock to allow near-lifetime-limited probe times and correspondingly reduced statistical uncertainties \cite{peik_laser_2006, leroux_-line_2017}.

The paper is organized as follows. In Section \ref{TASS} we show how robustness to QPS and TASS can be achieved by the application of a continuous detuned driving field. In Section \ref{SCHEME} we present the general CDD scheme for the construction of a robust optical clock transition. Then, in Section \ref{CRYSTAL} we consider the implementation of the scheme in the case of a multi-ion crystal clock and analyze the performance of the scheme in terms of the expected statistical uncertainties of the robust optical clock transition. We end with the conclusions in Section \ref{CONCLUSIONS}.

\section{Robustness to tensor shifts}
\label{TASS}
In the absence of hyperfine structure, tensor shifts including the quadrupole shift are proportional to $Q_{J,m_j}=J(J+1)-3m_{j}^{2}$ \cite{itano_external-field_2000}, where $J$ is the total angular momentum and $m_J$ the magnetic quantum number. These shifts reduce the precision of atomic clocks when not suppressed by suitable averaging schemes. Previously employed schemes to suppress the quadrupole shift include averaging the transition frequency over all $m_J$ states, since $\sum_{m_J=-J}^{J} Q_{J,m_j}=0$ \cite{dube_electric_2005}. Such an average will also eliminate the linear Zeeman shift, assuming the field does not change between the frequency measurements contributing to the average.
We propose a novel dynamical decoupling scheme in which robustness to these type of shifts is achieved by the application of a detuned driving field, mixing all $m_J$ states to form a dressed states with effective $Q_{J,m_j}=0$. While the cancellation scheme is general and applies to tensor shifts of arbitrary electronic states, we consider in the following the Hamiltonian of the $D_{5/2}$ states of e.g.~\Ca ions,  
\begin{equation}
H=g_d \mu_B B S_{z}+g_d\Omega_1\cos\left[(g_d \mu_B B-\delta_1)t\right]S_{x},
\end{equation}
where $g_d \mu_B B$ is the Zeeman splitting due to the static magnetic field $B$,  $g_d$ is the gyromagnetic ratio of the $D_{5/2}$ states, $\Omega_1$ is the Rabi frequency of the driving field, and $\delta_1$ is the detuning. Moving to the interaction picture (IP) with respect to $H_{0}=(g_d \mu_B B-\delta_1)S_{z}$ and taking the rotating-wave-approximation (RWA) ($(g_d \mu_B B-\delta_1)\gg\Omega_1$), results in 
\begin{equation}
H_{I}=\delta_1 S_{z}+\frac{g_d\Omega_1}{2}S_{x}.
\label{HI_detuned}
\end{equation}
Since the bare $D_{5/2}$ states have tensor shifts which are proportional to $Q_{5/2,\pm5/2}=-10$, $Q_{5/2,\pm3/2}=+2$, and $Q_{5/2,\pm1/2}=+8$, the tensor shifts of the dressed states (the eigenstates of $H_{I}$) are proportional to 
\begin{equation}
Q = 5 q(\delta, \Omega_1),\quad -q(\delta, \Omega_1), \quad -4 q(\delta, \Omega_1),
\label{Qdressed}
\end{equation}
with
\begin{equation}
q(\delta_1\Omega_1)=\frac{-8\delta_1^{2}+\left(g_d\Omega_{1}\right)^{2}}{4\delta_1^{2}+\left(g_d\Omega_{1}\right)^{2}}.
\end{equation}

\begin{figure}[t]
\centering{}\includegraphics[width=1\linewidth, right]{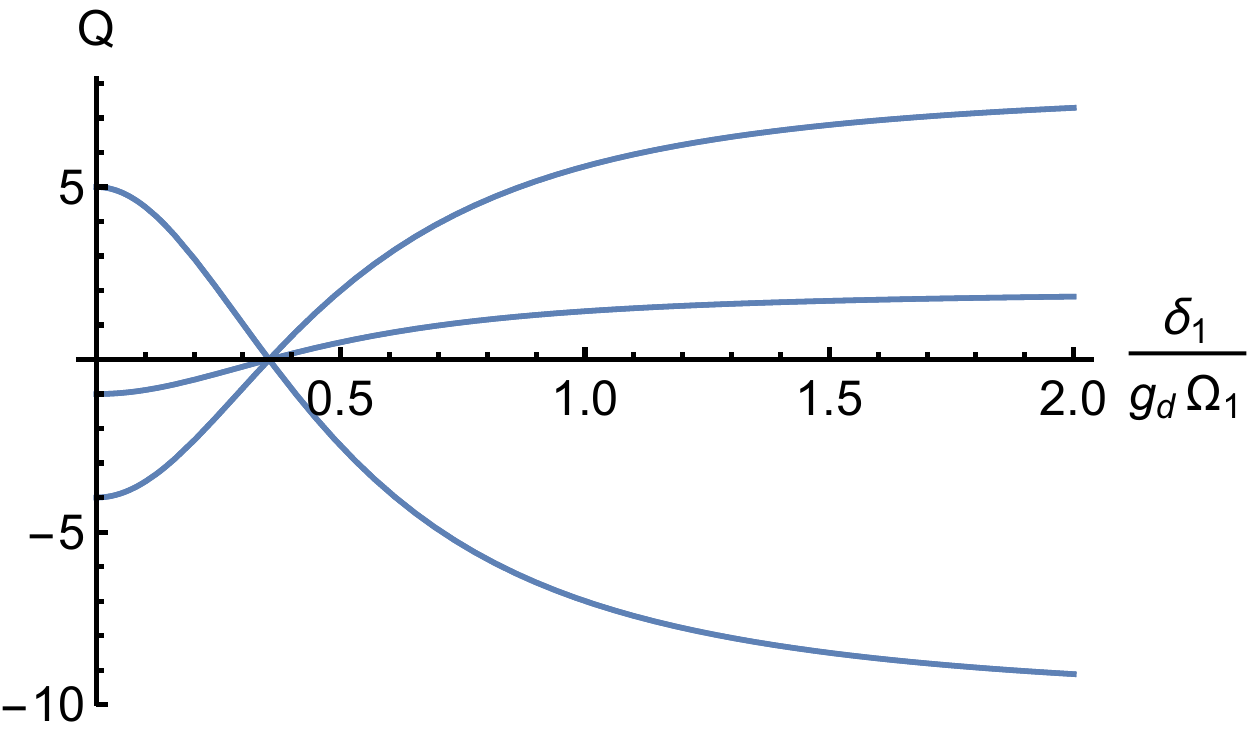}\protect\caption{The tensor shift factors $Q$ of the dressed states (Eq.~\ref{Qdressed}) as function of the detuning $\delta_1$. The tensor shifts vanish for $\delta_1=\pm\sqrt{\frac{1}{8}}\left(g_d\Omega_1\right)$.} 
\label{Qshift}
\end{figure}

Hence, by choosing $\delta_1=\pm\sqrt{\frac{1}{8}}\left(g_d\Omega_1\right)$, all of the dressed states have a zero (first order) tensor shift, $Q=0$ (see Fig.~\ref{Qshift}). This can also be understood in the lab frame, in which the coupling $\Omega_1$ drives a rotation among the bare states, averaging their shifts in time just as in the $m_J$ averaging scheme mentioned above. However, here the averaging takes place at a rate corresponding to the Rabi frequency rather than the experiment repetition rate, allowing it to suppress much faster field fluctuations.

The cancellation of the tensor shift to first order can also be understood as follows. The tensor and quadrupole shift operator $\hat{Q} = S^2-3 S_z^2$ can also be written as $\hat{Q} = S_x^2 + S_y^2 - 2 S_z^2$. Moving to a rotated basis (dressed states basis) defined by $z\rightarrow \cos(\theta)z + \sin(\theta)x$, $x\rightarrow \cos(\theta)x - \sin(\theta)z$, and $y\rightarrow y$, and neglecting purely off-diagonal terms we obtain that $\hat{Q}\approx S_z^2 \left(1-3\cos^2\left(\theta\right)\right)+S_x^2\left(1-3\sin^2\left(\theta\right)\right)+S_y^2$. To first order (diagonal terms) we have that $\hat{Q}\approx \left( S_+ S_- + S_- S_+ \right) \left(2-3 \sin^2\left(\theta\right)\right)+ S_z^2\left(1-3\cos^2\left(\theta\right)\right).$ Hence to first order, the tensor shift vanishes for $\cos\left(\theta\right)=\frac{1}{\sqrt{3}}$, which is analogous to magic-angle spinning in solid-state NMR spectroscopy \cite{hennel_mas_nmr_2005}.  
The detuned driving field results in dressed states, which are the eigenstates of Eq. \ref{HI_detuned}. The rotation to the dressed states basis is given by $U=e^{i \theta S_y}$,  where $\cos\left(\theta\right)=\frac{\delta_1^2}{\sqrt{\delta_1^2 + \frac{\left(g_d\Omega_{1}\right)^2}{4}}}$. For $\delta_1=\sqrt{\frac{1}{8}} \left(g_d\Omega_{1}\right)$ we obtain  $\cos\left(\theta\right)=\frac{1}{\sqrt{3}}$.

\section{The scheme}
\label{SCHEME}
\begin{figure}[t]
\centering{}\includegraphics[width=1\linewidth,right]{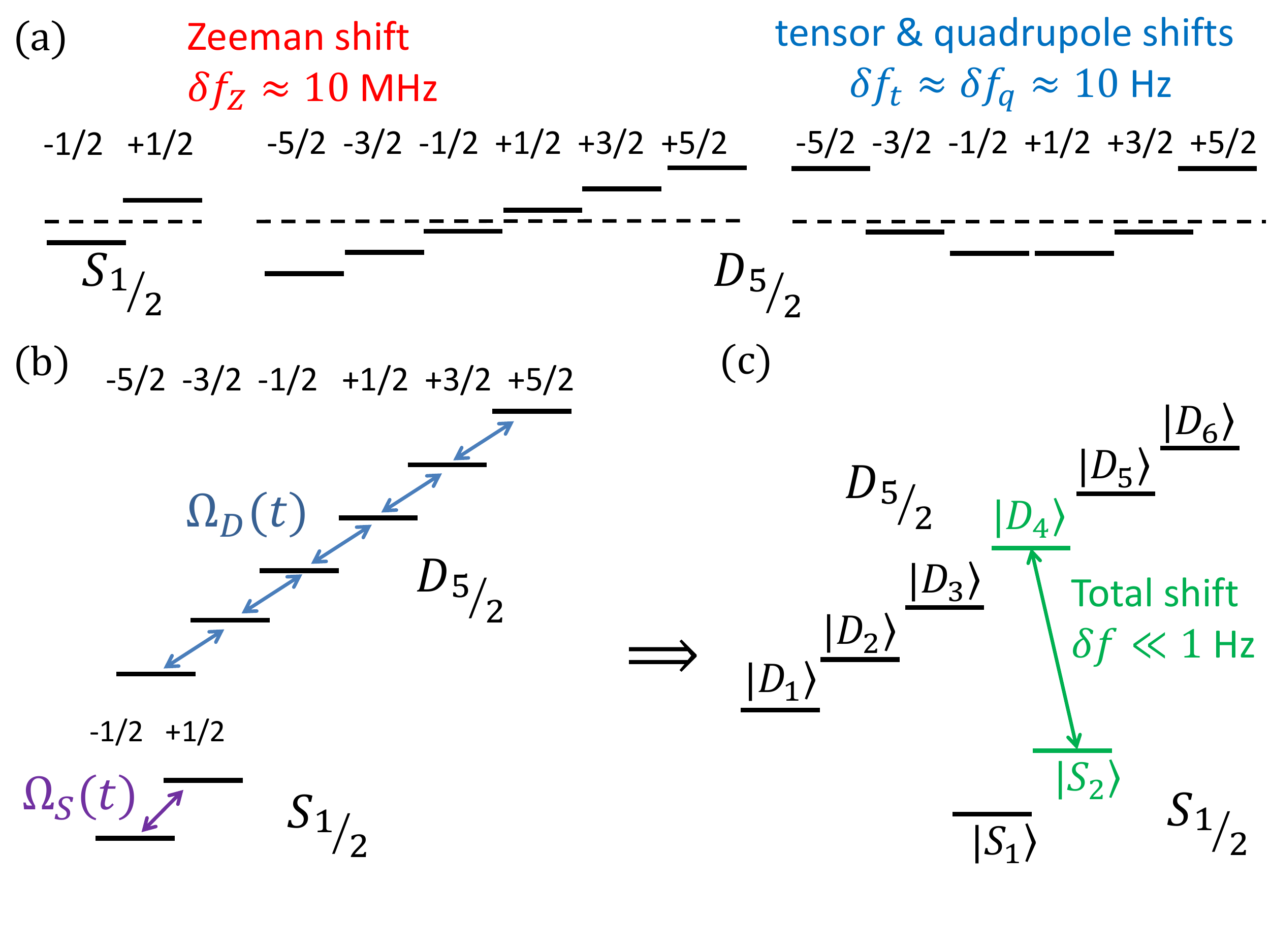}\protect\caption{Robust multi-ion crystal clock. (a) Typical Zeeman, tensor and quadrupole shifts of the unperturbed clock transition. (b) The applied driving fields of the scheme. (c) The double-dressed states. A robust optical transition with a total shift $\ll 1$~Hz and inhomogeneous linewidth $\sim 1$~Hz is constructed between the $\ket{S_2}$ and $\ket{D_4}$ states. }\label{Scheme}
\end{figure}

Our scheme is based on the application of continuous driving fields for the construction of a robust optical clock transition. We consider the sub-levels of the $S_{1/2}$ and $D_{5/2}$ states, the usual clock states in \Ca optical clocks. Four driving fields are employed, where two driving fields operate on the $S_{1/2}$ states, and two driving fields operate on the $D_{5/2}$ states. The first driving field of the $D_{5/2}$ states mitigates tensor shifts, including the QPS, as shown in the previous section. However, the $D_{5/2}$ states, as well as the  $S_{1/2}$ states, are still sensitive to the linear Zeeman shift. Moreover, the $D_{5/2}$ states are also sensitive to amplitude fluctuations of the driving field. The purpose of adding three more driving fields is to have enough control degrees of freedom that can be tuned such that the suppression of both  linear Zeeman shift and amplitude fluctuations of a single (dressed) \dsohddfh  transition is achieved.   
Hence, the driving scheme results in doubly-dressed  $S_{1/2}$ and $D_{5/2}$ states where one $S_{1/2} \leftrightarrow D_{5/2}$ transition (between the doubly-dressed states) is robust to magnetic shifts, quadrupole shifts, tensor shifts, and driving amplitude shifts (see Fig. \ref{Scheme}). For the $D_{5/2}$ ($S_{1/2}$) states, the Rabi frequencies, $\Omega_k$, and the detunings, $\delta_k$, of the first and second driving fields are denoted by $\{\Omega_1,\delta_1\}$ and $\{\Omega_2,\delta_2\}$ ($\{\Omega_3,\delta_3\}$ and $\{\Omega_4,\delta_4\}$), respectively. Similar to the derivation in Section \ref{TASS}, by moving to the first IP with respect to the frequencies of the first driving fields we obtain the dressed states, which are the eigenstates of 
\begin{equation}
H_I = \delta_1 S_z + \frac{g_d \Omega_1}{2} S_x + \delta_3 s_z + \frac{g_s \Omega_3}{2} s_x,
\end{equation}
where $S_i$ and $s_i$ are the spin matrices of the $D_{5/2}$ and $S_{1/2}$ states and  $g_d=6/5$ and $g_s=2$ are the gyromagnetic ratios of the $D_{5/2}$ and $S_{1/2}$ states, respectively. We proceed by moving to the basis of the dressed states and then to the IP with respect to the frequencies of the second driving fields and obtain the doubly-dressed states, which are the eigenstates of
\begin{equation}
H_{II} = \delta_2 S_z + \frac{g_d \Omega_2}{4} S_y + \delta_4 s_z + \frac{g_s \Omega_4}{4} s_y.
\label{HII} 
\end{equation}
A detailed derivation is given in Appendix I.

We consider the doubly-dressed $S_{1/2}$ and $D_{5/2}$ states with the smallest positive eigenvalue as the robust optical clock states. By setting  $\delta_1=\sqrt{\frac{1}{8}}\left(g_d\Omega_1\right)$ robustness to quadrupole and tensor shifts is attained. We denote by $\delta b$ and $\delta$ a magnetic shift and as the relative driving amplitude shift respectively. In order to achieve robustness to shifts in the magnetic and driving fields, we first add to the Hamiltonian the magnetic shift terms $\delta b S_z + \delta b s_z$, and driving shifts terms by $\Omega_k \rightarrow (1+\delta)\Omega_k$. We use a single parameter $\delta$ to describe the field amplitude fluctuations, which we assume to be correlated across all four dressing fields. This describes the experimental situation where the dominant variations in dressing-field amplitude are due to spatial inhomogeneities or to a common rf amplifier through which all four signals pass. The Hamiltonian of the double-dressed states $H_{II}$ (Eq.~\ref{HII}) now includes both $\delta b$ and $\delta$, which modify the eigenvalues (the energies) of the double-dressed states, and hence the optical transition frequency. We then calculate the power series expansion of the eigenvalues  to orders of $\delta b^i$ and $\delta^i$. We denote the series expansion terms of the magnetic shift of the $S_{1/2}$ and $D_{5/2}$ states by $Z_{S_i}\delta b^i$ and $Z_{D_i}\delta b^i$ respectively. Similarly, we denote the series expansion terms of the amplitude driving shift of the $S_{1/2}$ and $D_{5/2}$ states by $O_{S_i}\delta^i$ and $O_{D_i}\delta ^i$ respectively. The expansion terms of the correlated shifts are denoted by $ZO_{S_i}\delta b^i \delta$ and $ZO_{D_i}\delta b^i \delta$ for the $S_{1/2}$ and $D_{5/2}$ states respectively. We calculate the magnetic energy shifts, $Z_{S_i}\delta b^i$ and $Z_{D_i}\delta b^i$, up to fourth order ($i=1,...,4$),  driving energy shifts,  $O_{S_i}\delta^i$ and $O_{D_i}\delta^i$, up to second order ($i=1,2$), and the correlated shifts,  $ZO_{S_i}\delta b^i \delta$ and $ZO_{D_i}\delta b^i \delta$ ($i=1,2$) as function of the driving parameters, $\Omega_i$ and $\delta_i$.
A detailed derivation is given in Appendix II. We continue by defining a goal function, 
\begin{eqnarray}\label{eq:goal}
G &=& \sum_{i=1,j=1}^{i=4,j=2} |Z_{S_i}-Z_{D_i}|\delta b^i + |O_{S_j}-O_{D_j}|\delta^j \nonumber\\
&+& |ZO_{S_j}-ZO_{D_j}|\delta b^j \delta. 
\end{eqnarray}
Given distributions of the shifts, $\delta b$ and $\delta$, of a specific experimental set-up, we then define an averaged goal function, 

\begin{equation}\label{eq:goal_ave}
G_A = \langle G\left(\delta b_m, \delta_n\right) \rangle_{m,n},
\end{equation}
where $\delta b_m$ and $\delta_n$ are chosen randomly according to the given distributions, and numerically minimize $G_A$ over the driving parameters, $\Omega_k$ and $\delta_k$.  The numerical minimization results in optimal sets of values of the driving fields, for which robustness to shifts in the magnetic and driving fields is obtained. 

In the derivation of the optimal driving parameters we assumed the RWA and neglected cross-driving effects (for example, the effect of the $S_{1/2}$ driving field on the $D_{5/2}$ states). In the numerical analysis of the resulting shift distribution we took the counter-rotating terms of the driving fields and the cross-driving effect into account. In addition, the simulations where performed using the full driving Hamiltonian without making any approximations. For more details see  Appendix sections \ref{ME_sec}, \ref{BS_sec}, and \ref{numeric}.

\section{Robust Multi-Ion Crystal Clock}
\label{CRYSTAL}
We now consider the implementation of a robust multi-ion clock employing the proposed scheme and realizable with current ion trap technology. We estimate the dominating field inhomogeneities (QPS and TASS) for \Caf as a widely used clock species \cite{chwalla_absolute_2009,huang_comparison_2017,cao_compact_2017} with convenient properties, and discuss effects of micromotion.

As a trap platform, we consider a linear Paul trap, in which axial micromotion can be made sufficiently small \cite{herschbach_linear_2012, pyka_high-precision_2014}.  In large three-dimensional ion crystals, each ion will experience micromotion driven by the rf-field of the trap with an amplitude proportional to the distance from the trap's symmetry axis. Due to the construction of the trap, this micromotion is oriented along the radial axis (perpendicular to the trap axis). When probing along the radial degrees of freedom, the ions' coupling to the probe laser is diminished by the strong Doppler modulation of the laser in the ions' frame. We therefore choose to probe along the trap symmetry axis $z$ with a magnetic field oriented along the same direction.

The QPS of an ion in a multi-ion crystal is caused by electrical field gradients originating from the space charge of all other ions, and the electrical field gradients of the trap itself. The overall scale of the QPS is determined by the quadrupole moment expressed as a reduced matrix element, which for \Caf was measured to be $\Theta(3d,5/2) = 1.83(1)ea_0^2$ \cite{roos_designer_2006}. The QPS depends on the angle between the quantization axis and the electric field gradient as well as the state of the ion. The angle dependence is given by \cite{arnold_prospects_2015, itano_external-field_2000, roos_designer_2006} 

\begin{align}
f(\alpha,\beta,\gamma)& =  \frac{\partial E_{z}}{\partial z}\frac{1}{4}(3\cos(\beta)^2-1)\\ \nonumber
& + \frac{1}{2}\sin(2\beta)\left(\frac{\partial E_x}{\partial z}\cos\alpha  + \frac{\partial E_y}{\partial z}\sin\alpha\right)\nonumber \\
&+ \frac{1}{4}\sin^2\beta \left[\left(\frac{\partial E_x}{\partial x} - \frac{\partial E_y}{\partial y}\right)\cos(2\alpha) + 2\frac{\partial E_x}{\partial y}\sin(2\alpha)\right]\nonumber
\label{angledep}
\end{align}

with the Euler angles $\{\alpha,\beta,\gamma\}$ defined as in \cite{itano_external-field_2000}. The state dependence can be described by
\begin{equation}
 g(J,m_J) = \frac{J(J+1)-3m_J^2}{J(2J-1)},
 \end{equation}
with a total QPS of $\Delta f_\text{QPS}=\frac{\Theta}{h} \times f(\alpha,\beta,\gamma)\times g(J,m_J)$.

The Paul trap features two types of electrical field gradients: The rf-field employed for radial confinement and the static field gradient for axial confinement. The rf-field is averaged out and therefore does not lead to a quadrupole shift of the transition. The static gradient for axial trapping is constant and the same for all ions, causing no line broadening but a constant shift for each individual ion of the crystal. Provided the axial trap voltages are well controlled, this constant shift does not pose a limit on clock stability. Furthermore, this constant shift is cancelled by the dynamical decoupling scheme, see below.

The shift originating from the space charge will in general lead to inhomogeneous line broadening. The space charge-induced quadrupole shift falls off cubicly with distance and is therefore dominated by the ion's local environment. In a linear chain of ions, for instance, the quadrupole shift caused by the space charge results in a significant shift depending on the position of the ion, since contributions from the individual ions add up. For example, a chain of 30 \Caf ions exhibits an inhomogeneous shift of about 80~Hz across the chain (radial trap frequency $\omega_r = 2\pi\times 1 $~MHz and axial trap frequency $\omega_a = \omega_r/12$).  In a spherically symmetric crystal configuration, in contrast, the symmetry suppresses the quadrupole shift due to space charge to a large extent. In the limit of a crystal of infinite size a body-centered-cubic lattice forms and the quadrupole shift vanishes \cite{arnold_prospects_2015}.

%The shift originating from the space charge, in contrast, will in general lead to inhomogeneous line broadening. In a linear chain of ions, for instance, the quadrupole shift caused by the space charge results in a significant shift depending on the position of the ion, since contributions from the individual ions add up. For example, a chain of 30 \Caf ions exhibits an inhomogeneous shift  of about 80~Hz across the chain (radial trap frequency $\omega_r = 2\pi\times 1 $~MHz and axial trap frequency $\omega_a = \omega_r/12$). This shift can be strongly suppressed by a suitable choice of crystal geometry. In a spherically symmetric crystal configuration, for instance, the isotropic symmetry suppresses the quadrupole shift due to space charge to a large extent. In the limit of a crystal of infinite size, the quadrupole shift vanishes \cite{arnold_prospects_2015}. 

\begin{figure}[t]
\centering{}\includegraphics[width=1\linewidth]{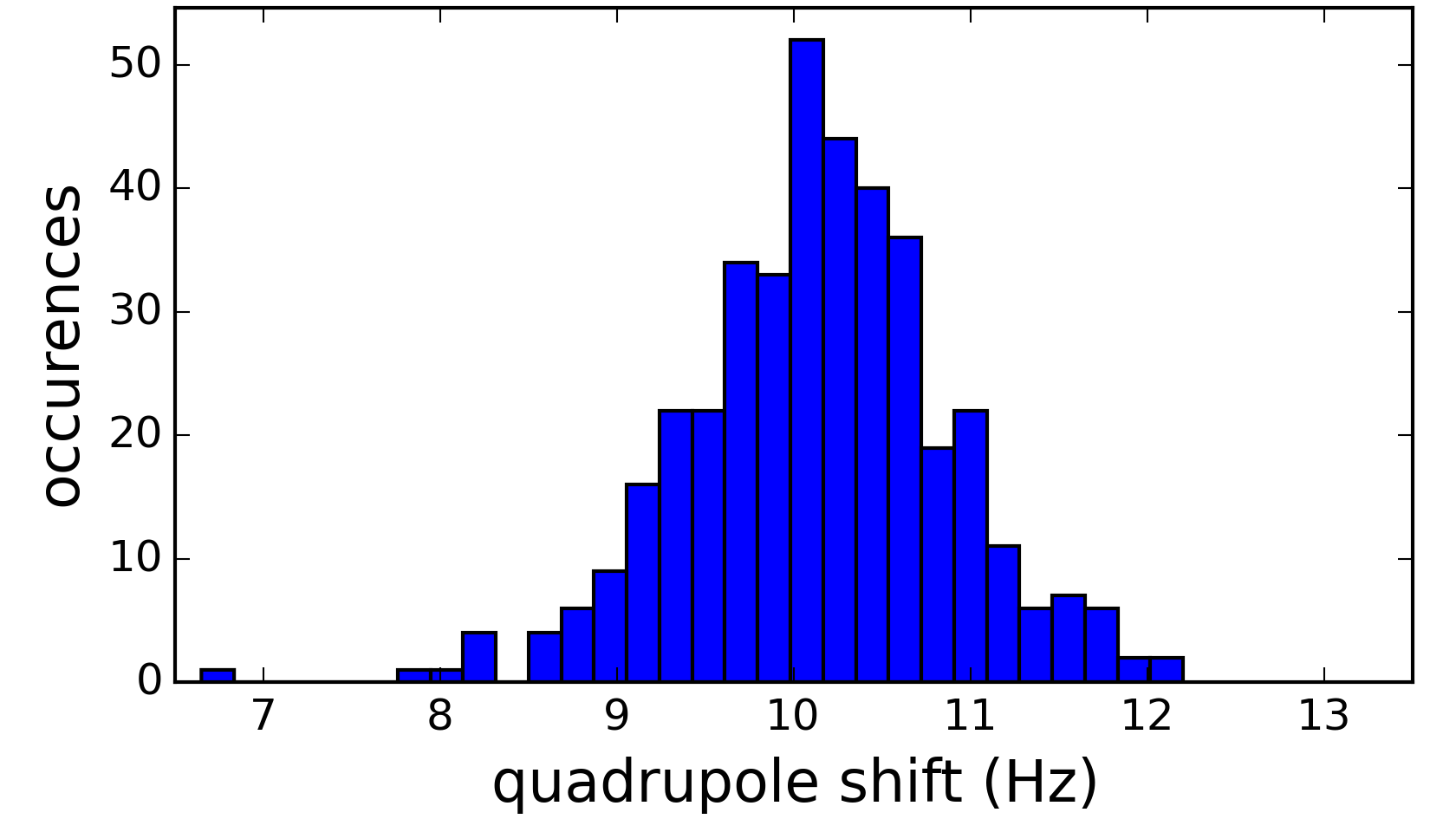}\protect\caption{Quadrupole shift for the $m_j =\pm \frac{5}{2}$ states for 400 ions in a linear Paul trap (spherical, isotropic crystal configuration). The distribution shows a standard deviation of 0.74~Hz and a static shift due to trap endcap voltage employed for axial confinement of about 10 Hz.}\label{quadshift}
\end{figure}

For this reason, we consider a spherical trap configuration in which it is straightforward to obtain near isotropic trap frequencies $\omega_{r,x}\approx \omega_{r,y}\approx \omega_a\approx 2\pi \times$1~MHz. A slight deviation from complete symmetry is desirable to pin the orientation of the crystal and allow efficient Doppler cooling. It is advantageous to work with as many ions as possible in order to improve the signal-to-noise ratio and hence the stability of the clock. On the other hand, cooling and trapping large crystals can be challenging and the size of the crystal needs to be reasonable so that the probe, cooling, and control lasers can address all the ions. Also, the probability of a background gas collision during clock interrogation grows with the number of ions. For our purpose, we consider a realistic implementation employing 400 \Caf ions, resulting in an isotropic crystal with approximately 30~$\mu$m radius for our proposed trap parameters. 

For this configuration, we estimate the QPS by first finding the equilibrium positions of the crystal ions by minimizing the pseudo-potential energy in the linear Paul trap. We next calculate the effective electric field tensor $\frac{\partial E_i }{\partial x_j}$ and the resulting quadrupole shift. In Fig.~\ref{quadshift} we show the resulting distribution of the QPS. The distribution shows a mean shift of about 10~Hz due to the static axial field gradient of the trap and a standard deviation of 0.74~Hz.

\begin{figure}[t]
\centering{}\includegraphics[width=1\linewidth]{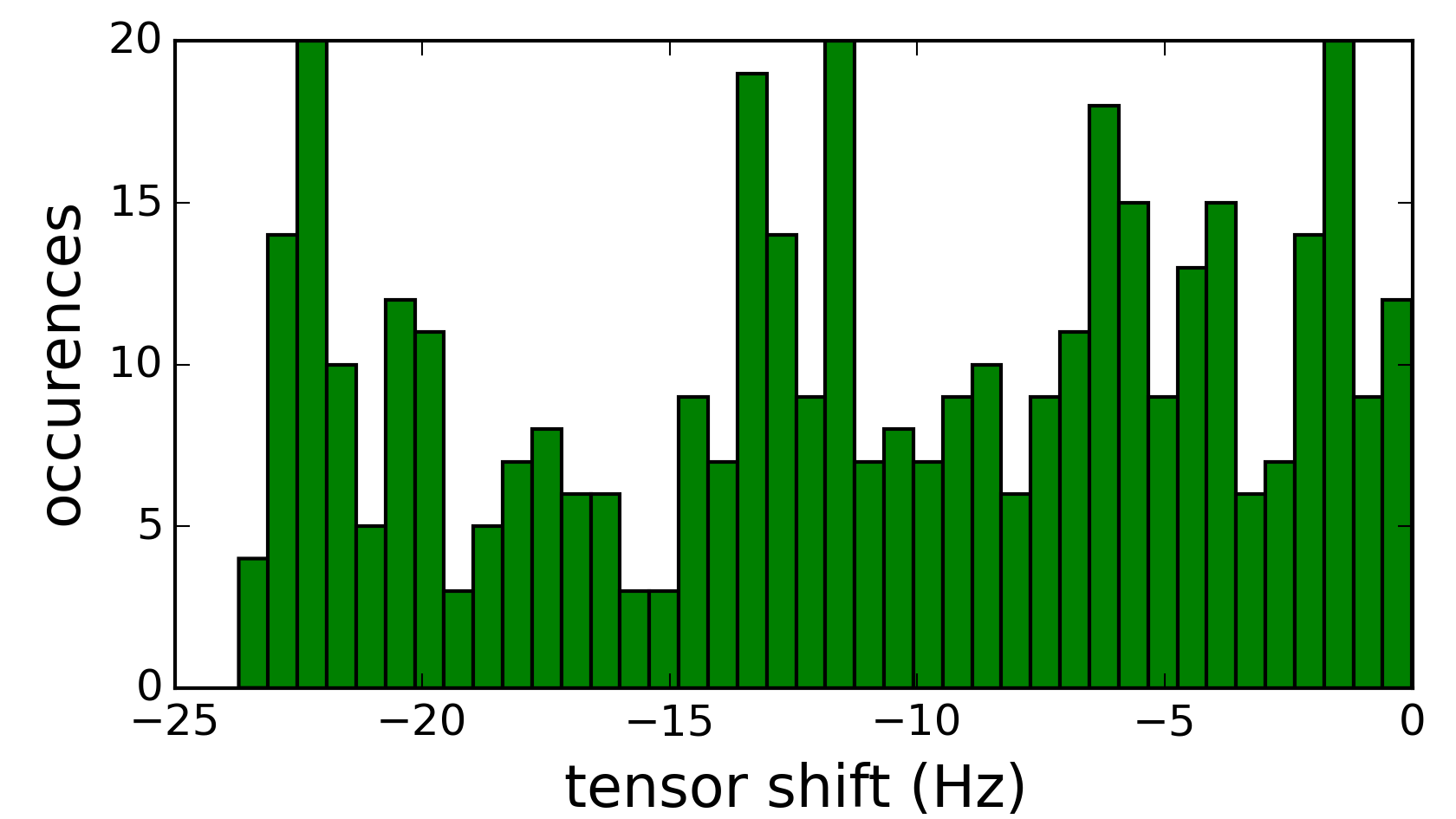}\protect\caption{Tensor shift for 400 ions for the $m_j =\pm \frac{5}{2}$ states in a linear Paul trap (spherical, isotropic crystal configuration). The shift distribution is calculated for a spherical, isotropic crystal configuration (for details see text). The tensor shift is approximately uniformly distributed with a width of $\sim 25$~Hz.}\label{tensorshift}
\end{figure}

In a spherical ion crystal, most of the ions are trapped away from the symmetry axis and the resulting rf-field at the individual ions' position causes a scalar and tensor ac Stark shift. While for \Caf  the scalar shift can be canceled by operating the trap at the ``magic'' drive frequency \cite{huang_frequency_2016}, the tensor ac Stark shift will result in shifted transition frequencies for each individual ion depending on its position. We estimate the TASS in our system by first calculating the amplitude of the rf-field due to the trap drive at each ion's position, employing a numerical simulation. The resulting TASS is then estimated by evaluating the time average $\langle ... \rangle$ over one cycle of the absolute value of the electrical field $E$ \cite{arnold_prospects_2015}:

\begin{equation}
\frac{\delta \nu}{\nu} = \frac{g(J,m_J)}{4}\frac{\alpha_{dc}}{h \nu}\langle E^2 \rangle,
\end{equation}
with the dc tensor polarizability $\alpha_{dc}$ from \cite{safronova_blackbody_2011}. Fig.~\ref{tensorshift} shows the result for our configuration, exhibiting an approximately uniform distribution with about 25~Hz width.

For the numerical optimization of the goal function Eq.~\ref{eq:goal}, we consider a  static magnetic field such that the Zeeman splitting  of the $S_{1/2}$ states is equal to $10$~MHz, where the uncertainty of the Zeeman splitting $g_s \delta b$ is normally distributed with a zero mean and a width of $1$~kHz. There is also a small contribution due to the second order Zeeman effect. For the assumed  magnetic field noise, this shifts amounts to an additional broadening of about 0.5~mHz (averaged over all relevant levels) and can therefore be neglected. In a real experiment, the rf driving fields used for dressing the states will be imperfect and show some fluctuations. We estimate the corresponding relative shifts of the driving fields $\delta$ to be normally distributed with a zero mean and a width of $ 4\times 10^{-4}$.

\begin{figure}[t]
\centering{}\includegraphics[width=1\linewidth,right]{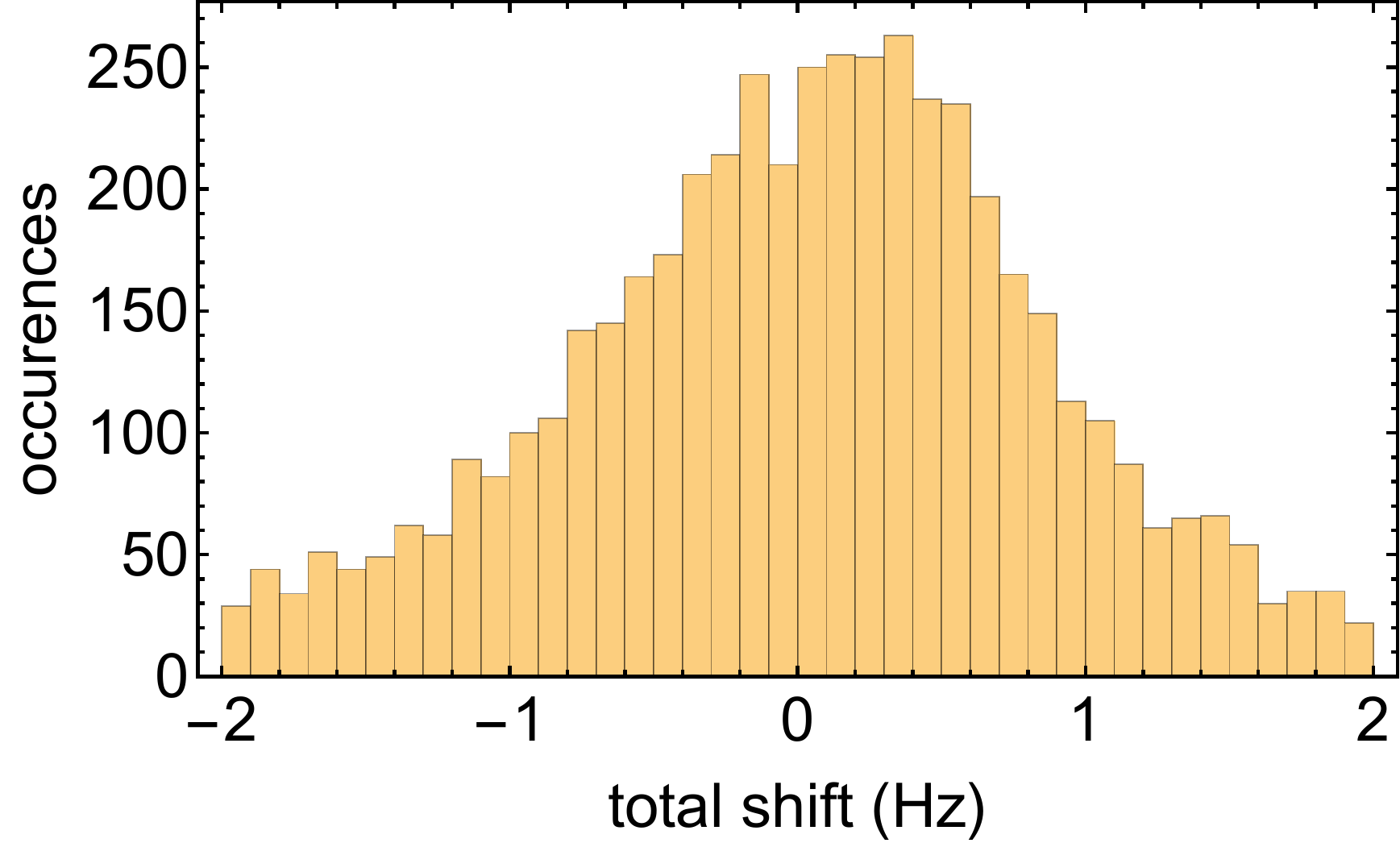}\protect\caption{Simulation results. Total shift distribution of the robust optical transition in the multi-ion spherical crystal configuration. $4927$ simulation trials where realized assuming the following distributions: (i) magnetic field uncertainty - the uncertainty of the $S_{1/2}$ Zeeman splitting is normally distributed with a zero mean and a width of $1$~kHz,  (ii) driving fields uncertainty - the relative drive amplitude uncertainty is normally distributed with a zero mean and a width of $0.4 \times 10^{-4}$, (iii) quadrupole shift - the quadrupole shift of the $m_j =\pm \frac{5}{2}$ states is normally distributed with a  mean of -10~Hz and a width of $1$~Hz, and (iv) tensor shift - the tensor shift of the $m_j =\pm \frac{5}{2}$ states is uniformly distributed between $-30$ and $0$~Hz. 
}\label{sim}
\end{figure}

In order to test the performance of the scheme in the multi-ion spherical crystal configuration, we have realized the averaged goal function $G_A$, which is given by Eq. \ref{eq:goal_ave}, with an average over $100$ pairs of $\{\delta b,\delta\}$ chosen randomly according to the above  distributions, and then numerically minimized $G_A$ to obtain optimal driving parameters (all in units of kHz): $\Omega_1=2\pi\times 225.3$, $\delta_1=2\pi\times 95.6$, $\Omega_2=2\pi\times 13.6$, $\delta_2=2\pi\times 5$, $\Omega_3=2\pi\times 93.6$, $\delta_3=2\pi\times 27.2$, and $\Omega_4=2\pi\times 14.8$, $\delta_4=2\pi\times 25.6$. Then, we simulated $4927$ trials of the robust clock-transition with these optimal driving parameters and the above distributions of the Zeeman, quadrupole, tensor and driving amplitude shifts.  In the simulations we took into account the effect of the $S_{1/2}$ $\left(D_{5/2}\right)$ drive on the $D_{5/2}$ $\left(S_{1/2}\right)$ states, as well as a correction to the counter-rotating terms of the driving fields (see Appendix Sections \ref{ME_sec}, \ref{BS_sec}, and \ref{numeric}). The simulation results, shown in Fig. (\ref{sim}), indicate that the shift distribution of the robust transition has a narrow width of $\sim 1$~Hz.

\begin{figure*}[t!]
\includegraphics[scale=0.32 ]{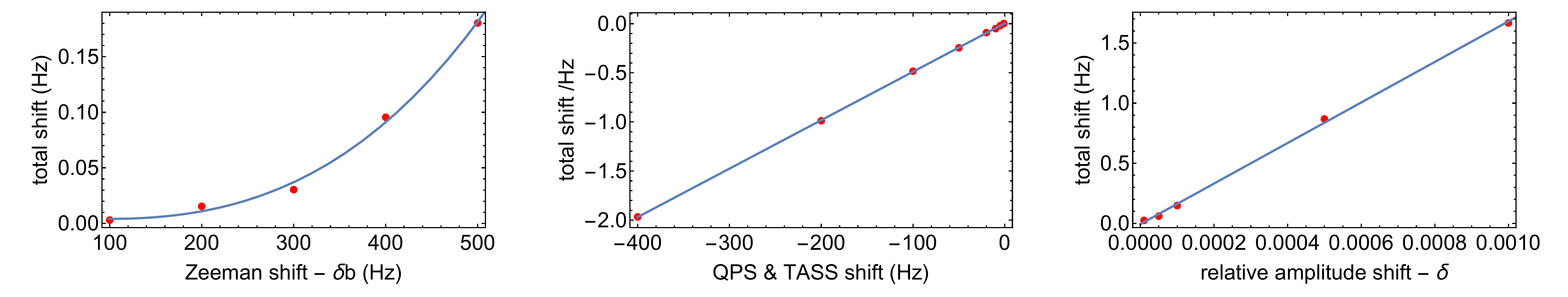}\caption{ Scaling of the robust transition shift. The shift of the robust transition as function of (i) the Zeeman shift of the $\ket{S_{1/2};+\frac{1}{2}}$ state (left), (ii) the quadrupole and tensor shifts of the the $\ket{D_{5/2};\pm\frac{5}{2}}$ states (middle), and (iii) the relative amplitude error of the driving fields.}
\label{scaling}
\end{figure*}

The distribution shown in the figure is slightly asymmetric and shifted away from zero.  The asymmetry is not a problem when using Ramsey interrogation with short broadband $\pi$/2 pulses which address all the ions at once, yielding a symmetric (cosine) central resonance.  However, the overall shift and the asymmetry lead to a probe-time dependent fractional bias in the clock frequency of around 0.119~Hz for a probe time of 150~ms, which would have to be evaluated and corrected in absolute frequency measurements.

This shift from the non-perturbed virtual (dressed) $\dsoh, m_J=0\leftrightarrow \ddfh,m_J=Q_{J,m_J}=0$ transition depends on the initial distribution of inhomogeneously shifted line centers and the shift suppression achievable with the selected dynamical decoupling parameters. A qualitative understanding of the influence of different parameters on this shift can be developed from simulation results for the scaling of the shift of the robust transition as a function of the magnitude of the individual shifts, namely, the Zeeman shift uncertainty, quadrupole and tensor shifts, and driving amplitude shift. The simulations shown in Fig.~\ref{scaling} were performed with the driving parameters used in the simulations of Fig.~\ref{sim}. For each individual shift, several simulations were conducted with different values of that individual shift while all other shift contributions were set to zero. 
When varying the magnetic field uncertainty, we see some higher-order (quadratic and cubic) contributions but no linear dependence, showing that the decoupling fields have greatly suppressed the first-order Zeeman shift.  This can be understood as an avoided crossing of the dressed levels with respect to the magnetic shift uncertainty $\delta b$.
The shift of the robust transition as function of the quadrupole and tensor shifts scales linearly. For a perfect drive of the $D_{5/2}$ states (assuming the RWA and no effect of the $S_{1/2}$ drive) we expect a quadratic scaling. However, the $S_{1/2}$ drive and the counter-rotating terms of the $D_{5/2}$ drive result in an amplitude mixing between the ideal dressed $D_{5/2}$ states, which reintroduces a first order contribution to the quadrupole and tensor shifts. The shift of the robust transition as a function of the drive amplitude error scales linearly as the robust transition frequency depends linearly on the drive amplitude. Note that the simulations were performed with fixed driving parameters, optimized as described above for a particular distribution of field shifts and drive amplitudes. For different assumptions on the distribution of field and amplitude fluctuations, different driving parameters could be derived that improve the scheme's performance case-by-case. Moreover, as in other CDD schemes, the performance of the scheme could, in principle, be improved by adding more driving fields. 

The coupling between the dressed $\dsoh$ states and the dressed $\ddfh$ states  is achieved via the laser coupling between the bare states that have a non-vanishing amplitude in the desired dressed states. Hence, the effective laser coupling strength is modified by the overlap between the bare states and the single or double dressed states.
%Because we consider double-dressed states, the effective laser coupling strength is further reduced by the overlap between the dressed states and the double-dressed states.
%since the coupling between double-dressed states is achieved via the coupling between the dressed states. 
In our case the laser coupling of the bare $\dsoh, m_J=+\frac{1}{2}\leftrightarrow \ddfh,m_J=+\frac{5}{2}$ transition is reduced by a  factor of $0.51$ $(0.3)$ for the transition between a single (double) dressed $\dsoh$ state and a single (double) dressed  $\ddfh$ state.
%, which is then reduced by a factor of $0.6$ for the transition between the double-dressed states of the robust optical transition. Hence, the overall reduction factor is equal to $0.3$. That is, the effective laser coupling strength between the double-dressed states of the robust optical transition is reduced by a factor of $0.3$ compared to the laser coupling strength of the bare states transition  $\dsoh, m_J=+\frac{1}{2}\leftrightarrow \ddfh,m_J=+\frac{5}{2}$. 
Note that the effective laser coupling strength should in any case be smaller than the energy gap of the double dressed states, which in our case is $\sim 6$ kHz.

\section{Conclusions}
\label{CONCLUSIONS}
We have proposed a CDD scheme that significantly suppresses the Zeeman shift as well as the quadrupole and tensor ac Stark frequency shifts of an optical clock transition for ion crystals containing on the order of 100 ions or more. We analyzed the proposed scheme in the case of a multi-ion crystal of 400 \Ca ions and showed that the shift of the robust transition is $\delta f \ll 1$~Hz with a width of $\sim 1$~Hz, which is close to the observed linewidth when probing the transition for a few hundred milliseconds. 
Using the results of \cite{leroux_-line_2017}, we estimate the achievable statistical uncertainty of a 400~ion \Ca clock when probed with a flicker floor-limited probe laser at $10^{-14}$ ($10^{-15}$) to be $5.8\times 10^{-16}/\sqrt{\tau/\text{s}}$ ($1.8\times 10^{-16}/\sqrt{\tau/\text{s}}$). This represents an order of magnitude improvement in instability over current single ion clocks \cite{huntemann_single-ion_2016, chou_frequency_2010, dube_$88mathrmsr+$_2015}, corresponding to a reduction in averaging time by a factor of 100. Through appropriate characterization of the residual line center shift away from an effective $\dsoh, m_J=0\leftrightarrow \ddfh, m_J=0$ transition, it is conceivable to not only obtain a reference with small statistical uncertainty, but also a low systematic uncertainty.

In our analysis we assumed that all driving fields suffer from driving fluctuations. Generating the second driving fields ($\Omega_2$ and $\Omega_4$) by a phase modulation, as proposed in \cite{cohen_continuous_2017}, should result in stable driving fields with negligible amplitude fluctuations. In this case we expect a further  improvement in the performance of the scheme.

One of the many potential applications of the proposed scheme is a cascaded clock \cite{rosenband_exponential_2013,borregaard_efficient_2013, hume_probing_2016} in which a clock laser is first stabilized to an ensemble of \Ca ions to improve its phase coherence time and thus allow extended probing times~\cite{peik_laser_2006, leroux_-line_2017} in a high-accuracy single-ion clock (e.g. \Al or \Yb). To bridge the difference in clock transition frequencies, a transfer scheme using a frequency comb would be employed \cite{stenger_ultraprecise_2002, scharnhorst_high-bandwidth_2015, hume_probing_2016}. Table~\ref{tab:instabilities} shows the achievable instabilities of an \Al clock for different initial clock laser instabilities. We have used the results from \citet{leroux_-line_2017} to determine the optimum probe times assuming flicker-floor limited laser instability and neglecting spontaneous emission from the excited clock state. Note that after stabilization to the \Ca ion crystal the laser exhibits a white frequency noise spectrum. As expected, the reduction in required averaging time is largest when the initial laser instability is large. As the laser improves, it approaches the quantum projection noise limit of the 400~\Ca ions and the gain is reduced. Even higher gain can be obtained from larger crystals or reference atoms with narrower linewidth than \Ca, such as lutetium \cite{arnold_prospects_2015, paez_atomic_2016}. The cascaded clock scheme enables short averaging times with lasers that are commercially available. Furthermore, many applications in fundamental physics, navigation and industry do not require ultimate accuracy \cite{poli_optical_2013,ludlow_optical_2015}, but rather high stability as provided by a dynamically-decoupled Coulomb crystal clock. We would like to note that during the preparation of this manuscript we became aware of a related independent work by Shaniv et al. \cite{shaniv_quadrupole_2018}.\\

\onecolumngrid
\begin{table*}
\begin{tabular}{|c|c|c|c|c|c|c|c|}
\hline 
$\sigma_l$ & $T_\mathrm{Ca}$ (s) & $\sigma(\tau)_\mathrm{Ca} \sqrt{\tau/\mathrm{s}}$ & $T_\mathrm{Al}$ (s) & $\sigma(\tau)_\mathrm{Al} \sqrt{\tau/\mathrm{s}}$ & $T_\mathrm{CaAl}$ (s) & $\sigma(\tau)_\mathrm{CaAl} \sqrt{\tau/\mathrm{s}}$ & $G$\\ 
\hline 
$10^{-14}$& 0.015 & $5.8\times 10^{-16}$ & 0.0035 & $3.6\times 10^{-15}$ & 0.25 & $5.1\times 10^{-16}$ & 50 \\ 
\hline 
$10^{-15}$& 0.15 & $1.8\times 10^{-16}$ & 0.035 & $1.1\times 10^{-15}$ & 0.78 & $2.9\times 10^{-16}$ & 14\\ 
%\hline 
%$10^{-16}$& 1.5$^*$ & $5.8\times 10^{-17*}$ & 0.35 & $3.6\times 10^{-16}$ & 2.5$^*$ & $1.6\times 10^{-16*}$ & 5\\ 
\hline 
\end{tabular} 
\caption{Estimated statistical uncertainties. The flicker-floor limited performance of the clock laser is denoted by $\sigma_l$, which is assumed to be independent of the averaging time. The Allan deviations and optimum probe times for different configurations $k$ are denoted by $\sigma_k$ and $T_k$, respectively. The investigated systems are 400~\Ca ions using the described DD scheme ($k=\mathrm{Ca}$), a single \Al ion ($k=\mathrm{Al}$), and a cascaded scheme in which a single \Al ion is probed by a laser pre-stabilized through a cloud of 400~\Ca ions ($k=\mathrm{CaAl}$). The reduction in averaging time to achieve a certain statistical measurement uncertainty is given by $G$. 
}
\label{tab:instabilities}

\end{table*}
\twocolumngrid

\section{Acknowledgements}
We thank M. Barrett for stimulating discussions.
We acknowledge support from the DFG through CRC 1128 (geo-Q), project A03 and CRC 1227 (DQ-mat), project B03. This joint research project was financally supported by the State of Lower-Saxony, Hannover, Germany. \\

\onecolumngrid
\appendix
\newpage
\section{The dressed states}
In this section we present the construction of the (ideal) dressed states. By ideal we mean that there are no noise, uncertainties, or systematic shifts, the rotating-wave-approximation (RWA) is valid, and that the driving fields of the $S_{1/2}$ $\left(D_{5/2}\right)$ do not operate on the $D_{5/2}$ $\left(S_{1/2}\right)$ states.
\subsection{The $D_{5/2}$ states}
The driving Hamiltonian of the $D_{5/2}$ states is given by
\begin{equation}
H_D= g_d \mu_B B S_z + g_d \Omega_1 \cos \left[ \left(g_d \mu_B B - \delta_1\right)t \right]  S_x  + g_d\Omega_2 \cos \left[ \left(g_d \mu_B B - \delta_1\right)t + \frac{\pi}{2}\right]\cos \left[ \left(\sqrt{\left(\frac{g_d \Omega_1}{2}\right)^2+\delta_1^2} - \delta_2\right)t \right]  S_x, 
\end{equation} 
where $g_d \mu_B B$ is the Zeeman splitting due to the static magnetic field $B$, $g_d=6/5$ is the gyromagnetic ratio of the $D_{5/2}$ states, $S_z$ and $S_x$ are the  $z$ and $x$ spin-$5/2$ matrices, $\Omega_1$, $\Omega_2$, $\delta_1=\sqrt{\frac{1}{8}} g_d \Omega_1 $ and $\delta_2$ are the Rabi frequencies and the detunings of the driving fields, respectively. Moving to the interaction picture (IP) with respect to the first drive ($\Omega_1$) with $H_{01} = \left(g_d \mu_B B -\delta_1\right) S_z$ and assuming the RWA $\left(g_d \mu_B B -\delta_1\gg \Omega_1\right)$ we get
\begin{equation}
H_D^{I_1}= \delta_1 S_z +\frac{g_d \Omega_1}{2} S_x  + \frac{g_d\Omega_2}{2}\cos \left[ \left(\sqrt{\left(\frac{g_d \Omega_1}{2}\right)^2+\delta_1^2} - \delta_2\right)t \right]  S_y. 
\end{equation}   
We continue by moving to the basis of the dressed states with $U_1=e^{i \theta_{d}S_y}$, where $\theta_{d}=\arccos\left(\frac{\delta_1}{\sqrt{\delta_1^2+\left(\frac{g_d \Omega_1}{2}\right)^2}}\right)$, which leads to
\begin{equation}
H_D^{I_1}= \sqrt{\delta_1^2+\left(\frac{g_d \Omega_1}{2}\right)^2} S_z  + \frac{g_d\Omega_2}{2}\cos \left[ \left(\sqrt{\left(\frac{g_d \Omega_1}{2}\right)^2+\delta_1^2} - \delta_2\right)t \right]  S_y,
\end{equation}     
and then to the second IP with respect to $H_{02} = \left(\sqrt{\left(\frac{g_d \Omega_1}{2}\right)^2+\delta_1^2} - \delta_2\right) S_z$. Assuming the RWA, $\left(\sqrt{\left(\frac{g_d \Omega_1}{2}\right)^2+\delta_1^2} - \delta_2\right) \gg \Omega_2$, we obtain
\begin{equation}
H_S^{I_2}= \delta_2 S_z +\frac{g_d \Omega_2}{4} S_y. 
\end{equation}   
The eigenstates of $H_D^{I_2}$ are the double-dressed $D_{5/2}$ states.  The eigenstate with the smallest positive eigenvalue of $\frac{1}{2}\sqrt{\delta_2^2+\left(\frac{g_d\Omega_2}{4}\right)^2}$ is used for the robust optical transition. 
   
\subsection{The $S_{1/2}$ states}

The driving Hamiltonian of the $S_{1/2}$ states is given by
\begin{equation}
H_S= g_s \mu_B B s_z + g_s \Omega_3 \cos \left[ \left(g_s \mu_B B - \delta_3\right)t \right]  s_x  + g_s\Omega_4 \cos \left[ \left(g_s \mu_B B - \delta_3\right)t + \frac{\pi}{2}\right]\cos \left[ \left(\sqrt{\left(\frac{g_s \Omega_3}{2}\right)^2+\delta_3^2} - \delta_4\right)t \right]  s_x, 
\label{Hs}
\end{equation} 
where $g_s \mu_B B$ is the Zeeman splitting due to the static magnetic field $B$, $g_s=2$ is the gyromagnetic ratio of the $S_{1/2}$ states, $s_z$ and $s_x$ are the  $z$ and $x$ spin-$1/2$ matrices, $\Omega_3$, $\Omega_4$, $\delta_3$ and $\delta_4$ are the Rabi frequencies and the detunings of the driving fields, respectively. Moving to the interaction picture (IP) with respect to the first drive ($\Omega_3$) with $H_{01} = \left(g_s \mu_B B -\delta_3\right) s_z$ and assuming the RWA $\left(g_s \mu_B B -\delta_3\gg \Omega_3\right)$ we get
\begin{equation}
H_S^{I_1}= \delta_3 s_z +\frac{g_s \Omega_3}{2} s_x  + \frac{g_s\Omega_4}{2}\cos \left[ \left(\sqrt{\left(\frac{g_s \Omega_3}{2}\right)^2+\delta_3^2} - \delta_4\right)t \right]  s_y. 
\end{equation}   
We continue by moving to the basis of the dressed states with $U_1=e^{i \theta_{s} s_y}$, where $\theta_{s}=\arccos\left(\frac{\delta_3}{\sqrt{\delta_3^2+\left(\frac{g_s \Omega_3}{2}\right)^2}}\right)$, which leads to
\begin{equation}
H_S^{I_1}= \sqrt{\delta_3^2+\left(\frac{g_s \Omega_3}{2}\right)^2} s_z  + \frac{g_s\Omega_4}{2}\cos \left[ \left(\sqrt{\left(\frac{g_s \Omega_3}{2}\right)^2+\delta_3^2} - \delta_4\right)t \right]  s_y,
\end{equation}     
and then to the second IP with respect to $H_{02} = \left(\sqrt{\delta_3^2+\left(\frac{g_s \Omega_3}{2}\right)^2} -\delta_4\right) s_z$. Assuming the RWA, $\left(\sqrt{\delta_3^2+\left(\frac{g_s \Omega_3}{2}\right)^2} -\delta_4\right) \gg \Omega_4$, we obtain
\begin{equation}
H_S^{I_2}= \delta_4 s_z +\frac{g_s \Omega_4}{4} s_y. 
\end{equation}   
The eigenstates of $H_S^{I_2}$ are the double-dressed $S_{1/2}$ states.  The eigenstate with the positive eigenvalue of $\frac{1}{2}\sqrt{\delta_4^2+\left(\frac{g_s\Omega_4}{4}\right)^2}$ is used for the robust optical transition.

\section{Magnetic and Drive shifts}
\label{shift_sec}
In this section we show how the expansions of the magnetic shifts, $Z_{S_i}$ and $Z_{D_i}$, the drive shifts, $O_{S_i}$ and $O_{D_i}$, and the correlated shifts, $ZO_{S_i}$ and $ZO_{D_i}$ are derived. 
In the derivation we assumed the RWA and neglected cross-driving effects. In the numerical simulations we took the counter-rotating terms of the driving fields and the cross-driving effect into account (see  Appendix Sections \ref{ME_sec},\ref{BS_sec}). The simulations were performed using the full driving Hamiltonian without making any approximations (see Appendix Section \ref{numeric}). 
For simplicity we'll show the derivation for the $S_{1/2}$ states. The derivation for the $D_{5/2}$ follows the same calculations. 

We start by adding to the driving Hamiltonian of the $S_{1/2}$ states, Eq. (\ref{Hs}), a magnetic noise term, which is given by $g_s \delta b s_z$. The drive shift is introduced by replacing $\Omega_3$ and $\Omega_4$ by $\Omega_3\left(1+\delta\right)$ and $\Omega_4\left(1+\delta\right)$, where $\delta$ represents a relative error shift of the driving fields. 
We assume that the relative errors of the driving fields are correlated since we expect that these errors are mostly due to changes in the amplifier chain and antenna, which are common to all drives. Moving to the IP with respect to the first drive as before and assuming the RWA $\left(g_s \mu_B B -\delta_3\gg \Omega_3\right)$ we now obtain
\begin{equation}
H_S^{I_1}=g_s \delta b s_z +  \delta_3 s_z +\frac{g_s \Omega_3\left(1+\delta\right)}{2} s_x  + \frac{g_s\Omega_4\left(1+\delta\right)}{2}\cos \left[ \left(\sqrt{\left(\frac{g_s \Omega_3}{2}\right)^2+\delta_3^2} - \delta_4\right)t \right]  s_y. 
\end{equation}
We continue by moving to the basis of the dressed states, including the shifts, with $U_1=e^{i \theta_{s} s_y}$, where $\theta_{s}=\arccos\left(\frac{\delta_3 + g_s \delta b}{\sqrt{\left(\delta_3 + g_s \delta b\right)^2+\left(\Omega_3\left(1+\delta\right)\right)^2}}\right)$, which leads to
\begin{equation}
H_S^{I_1}= \sqrt{\left(\delta_3 + g_s \delta b\right)^2+\left(\Omega_3\left(1+\delta\right)\right)^2} s_z  + \frac{g_s\Omega_4\left(1+\delta\right)}{2}\cos \left[ \left(\sqrt{\left(\frac{g_s \Omega_3}{2}\right)^2+\delta_3^2} - \delta_4\right)t \right]  s_y,
\end{equation} 
and then to the second IP with respect to $H_{02} = \left(\sqrt{\delta_3^2+\Omega_3^2} -\delta_4\right) s_z$. Assuming the RWA, $\left(\sqrt{\delta_3^2+\Omega_3^2} -\delta_4\right) \gg \Omega_4$, we obtain
\begin{equation}
H_S^{I_2}= \left(\delta_4+\sqrt{\left(\delta_3 + g_s \delta b\right)^2+\left(\Omega_3\left(1+\delta\right)\right)^2} -\sqrt{\delta_3^2+\Omega_3^2}\right) s_z +\frac{g_s \Omega_4\left(1+\delta\right)}{4} s_y.  
\end{equation} 
The positive eigenvalue, which we consider for the robust clock transition, is given by
\begin{eqnarray}
e_s &=& \frac{1}{4} \left[4 \text{$\delta_4$} \left(2 \sqrt{(\delta +1)^2 \text{$\Omega_3$}^2+(\text{$\delta_3$}+2 \text{$\delta b$})^2}-2 \sqrt{\text{$\delta_3$}^2+\text{$\Omega_3
$}^2}+\text{$\delta_4$}\right)-8 \sqrt{\left(\text{$\delta_3$}^2+\text{$\Omega_3$}^2\right) \left((\delta +1)^2 \text{$\Omega_3$}^2+(\text{$\delta_3 $}+2 \text{$\delta b
   $})^2\right)}\right.\nonumber\\
   &+&4 \left.(\delta  (\delta +2)+2) \text{$\Omega_3$}^2+(\delta +1)^2 \text{$\Omega_4$}^2+8 \left(\text{$\delta_3$}^2+2 \text{$\delta_3$} \text{$\delta b$}+2
   \text{$\delta b$}^2\right)\right]^{\frac{1}{2}}.
\end{eqnarray}

Following the same calculations for the $D_{5/2}$ states we obtain the lowest positive eigenvalue of the double-dressed $D_{5/2}$ states,  which is given by
\begin{equation}
e_d = \frac{1}{2} \sqrt{\frac{1}{100} \left(3 \sqrt{2 \left(2 \delta ^2+4 \delta +3\right) \text{$\Omega_1$}^2+16 \text{$\delta b$}^2+8 \sqrt{2} \text{$\delta b$} \text{$\Omega_1
   $}}+10 \text{$\delta_2$}-3 \sqrt{6} \text{$\Omega_1$}\right)^2+\frac{9}{100} (\delta +1)^2 \text{$\Omega_2$}^2}.
\end{equation}     
The magnetic shifts, $Z_{S_i}$ and $Z_{D_i}$, the drive shifts, $O_{S_i}$ and $O_{D_i}$, and the correlated shifts, $ZO_{S_i}$ and $ZO_{D_i}$, are obtained by the power series expansion of $e_s$ and $e_d$ to orders of $\delta b^i$ and $\delta^i$.

\section{Modified energy gaps from cross-driving}
\label{ME_sec}
A drive of the $S_{1/2}$ $\left(D_{5/2}\right)$ states also drives the $D_{5/2}$ $\left(S_{1/2}\right)$ states off-resonantly and results in a Stark shift of the initial sub-levels energy gap.   For simplicity we'll show the derivation for the $S_{1/2}$ states. The derivation for the $D_{5/2}$ follows the same calculations. 
% Follow First_Drive_frequencies.nb and First_Drive_frequencies_short.nb for derivation (in folder \second version\Sim\Sim_1) 
Consider the off-resonant drive of the $S_{1/2}$ states, 
\begin{equation}
H_s = \omega_0 s_z + g_s \Omega \cos \left[ \left(\omega_0 -\delta\right) t \right] s_x,
\end{equation}
where $\omega_0=g_s \mu_B B$ is the Zeeman splitting and $\delta$ is the detuning of the drive. We first move the the IP of the counter-rotating terms of the drive with respect to $H_{01}= - \left(\omega_0 -\delta\right) s_z$. This results in
\begin{equation}
H_s^{I_1} = \left(2 \omega_0-\delta\right) s_z + \frac{g_s \Omega}{2} \left[ s_x + \frac{1}{2} \left(e^{i \left(2\omega_0 -\delta\right) t} \sigma_+ + e^{-i \left(2\omega_0 -\delta\right) t} \sigma_-\right)\right].
\end{equation}
We continue by moving to the diagonalized basis of the time-independent part of $H_s^{I_1}$ with  $U_1=e^{i \theta_{s} s_y}$, where \newline $\theta_{s}=\arccos\left(\frac{\left(2\omega_0 -\delta\right) }{\sqrt{\left(2\omega_0 -\delta\right) ^2+\left(\frac{g_s \Omega}{2}\right)^2}}\right)$, and then to a second IP of the rotating terms of the drive with respect to $H_{02}=  2\left(\omega_0 -\delta\right) s_z$. The time-independent part of $H_s^{I_2}$ is given by
\begin{equation}
H_s^{I_2} \approx \left(2\left(\delta-\omega_0\right) + \sqrt{\left(\delta-2 \omega_0\right)^2 + \left(\frac{g_s \Omega}{2}\right)^2} \right) s_z +  \frac{g_s \Omega}{4}\left(1+\frac{2 \omega_0 -\delta}{\sqrt{\left(2 \omega_0 -\delta\right)^2+ \left(\frac{g_s \Omega}{2}\right)^2 }}\right) s_x ,
\end{equation}
and hence, the eigenvalues are equal to $\pm \left[\left(\left(\delta-\omega_0\right) + \sqrt{\left(\delta-2 \omega_0\right)^2 + \left(\frac{g_s \Omega}{2}\right)^2} \right)^2+ \frac{g_s \Omega}{8}\left(1+\frac{2 \omega_0 -\delta}{\sqrt{\left(2 \omega_0 -\delta\right)^2+ \left(\frac{g_s \Omega}{2}\right)^2 }}\right)^2 \right]^{\frac{1}{2}}$, 
which gives the modified energy gap. Plugging in the parameters of the first $D_{5/2}$ drive, ($\Omega_1$ and $\delta=\omega_0-\omega_1$, where $\omega_1=g_d \mu_B B -\delta_1$ and $g_s=2$ we obtained the modified energy gap of the $S_{1/2}$ states, which is given by
\begin{equation}
E_S =\omega_1 + \sqrt{\frac{1}{4} \Omega_1^2 \left(\frac{\omega_0+\omega_1}{\sqrt{(\omega_0+\omega_1)^2+\Omega
_1^2}}+1\right)^2+\left(\sqrt{(\omega_0+\omega_1)^2+\Omega_1^2}-2\omega_1\right)^2}.
\end{equation}
 For the $D_{5/2}$ states the modified energy gap is given by 
 \begin{equation}
E_D = \omega_3-\frac{1}{5} \sqrt{9 \Omega_3^2 \left(\frac{5 (\omega_0+\omega_3)}{2 \sqrt{25 (\omega_0+\omega
_3)^2+9 \Omega_3^2}}+\frac{1}{2}\right)^2+\left(\sqrt{25 (\omega_0+\omega_3)^2+9 \Omega_3^2}-10 \omega
_3\right)^2},
\end{equation}
where here $\omega_0 = g_d \mu_B B$. 
 Expanding the modified energy gaps in a power series of $\Omega_1$ and $\Omega_3$ results in $E_S \approx \omega_0 + \frac{\omega_0 \Omega_1^2}{\omega_0^2-\omega_1^2}$ and $E_D \approx \omega_0 - \frac{9 \omega_0 \Omega_3^2}{25 \left(\omega_3^2-\omega_0^2\right)}$. Note that the second order correction could also be calculated by an effective Hamiltonian  approach.

\section{Correction of the Bloch-Siegert shift}
\label{BS_sec}
In this section we give a detailed derivation of the correction of the Bloch-Siegert shift. The correction can be understood as  follows. Without the correction, we first consider the dressed states due to the rotating-terms of a driving field and then consider the effect of the off-resonance counter-rotating terms on the dressed states. This results in an energy shift of the dressed states and  (a time-dependent) amplitude-mixing between the dressed states, which is detrimental to the scheme because it modifies the shifts of the dressed states considered for the robust transition. To correct this effect, we first consider the effect of the counter-rotating terms on the bare states, and then fix the frequency of the drive accordingly such that the rotating-terms will drive the modified bare states. Consider, for example, the on-resonance driving Hamiltonian
\begin{equation}
H_d= \frac{\Omega_1}{2} \sigma_z + \frac{\Omega_2}{2} \cos \left( \omega_2 t \right)  \sigma_x.
\end{equation} 
Instead of moving to the interaction picture (IP) of the rotating frame we first move to the IP of the counter-rotating frame with respect to $H_0 = -\frac{\Omega_1}{2} \sigma_x$ and obtain
\begin{equation}
H_I= \frac{\Omega_1+\omega_2}{2} \sigma_z +\frac{\Omega_2}{4}\sigma_z   +\frac{\Omega_2}{4}\left(\sigma_+ e^{-2i\omega_2 t}+\sigma_ e^{+2i\omega_2 t} \right).
\end{equation} 
We continue by moving to the diagonal basis of the time-independent part of $H_I$, 
\begin{equation}
H_I\approx \frac{1}{4} \sqrt{4(\Omega_1+\omega_2)^2+\Omega_2^2} \sigma_z  +\frac{\tilde{\Omega}_2}{2}\left( \sigma_+ e^{-2i\omega_2 t}+\sigma_ e^{+2i\omega_2 t} \right). 
\end{equation} 
If we choose $2\omega_2=\frac{1}{2} \sqrt{4(\Omega_1+\omega_2)^2+\Omega_2^2}$ the rotating terms are on-resonance with the energy gap of the modified bare states. The on-resonance condition is given by 
\begin{equation}
\omega_2=\frac{1}{6}\left(2 \Omega_1+ \sqrt{16 \Omega_1^2+3 \Omega_2^2}\right).
\end{equation}
In addition, due to the basis change from the basis of the bare states to the basis of the modified bare states, the Rabi frequency of the drive is slightly modified, $\Omega_2 \rightarrow \tilde{\Omega}_2$, where
\begin{equation}
\label{BSRabi}
\tilde{\Omega}_2 = \frac{1}{2} \Omega_2 \left(1+\frac{2 (\omega_2 +\Omega_1)}{\sqrt{4 (\omega_2 +\Omega_1)^2+\Omega_2^2}}\right)\approx\Omega_2-\frac{\Omega_2^3}{16 (\omega_2 +\Omega_1)^2}.
\end{equation}
Because $\tilde{\Omega}_2$ corresponds to an optimal driving parameter, we must take that into account and set the initial Rabi frequency, $\Omega_2$, accordingly.    

\section{Numerical analysis}
\label{numeric}  
In this section we provide a more detailed description of the numerical analysis of the scheme in the case of a robust multi-ion crystal clock that is presented in Section IV of the main text. 
\subsection{Optimized driving parameters}
For the optimization of the goal function (see Section III in main text and Appendix Section \ref{shift_sec}),
\begin{equation}
G = \sum_{i=1,j=1}^{i=4,j=2} |Z_{S_i}-Z_{D_i}|\delta b^i + |O_{S_j}-O_{D_j}|\delta^j + |ZO_{S_j}-ZO_{D_j}|\delta b^j \delta,
\end{equation} 
we assume that the magnetic shift uncertainty $\delta b$ is normally distributed with a zero mean and a width of $0.5$~kHz (so the width of the $S_{1/2}$ Zeeman splitting is $1$~kHz), $\delta b \sim \mathcal{N}\left(0,0.5\right)$~kHz, and that the relative drive shift $\delta$ is normally distributed with a zero mean and a width of $4 \times 10^{-4}$, $\delta \sim \mathcal{N}\left(0,4 \times 10^{-4}\right)$.
Given these distributions of $\delta b$ and $\delta$, we defined an averaged goal function, $G_A = \langle G\left(\delta b_m, \delta_n\right) \rangle_{m=1,n=1}^{m=100,n=100}$ over $100$ realizations of  $\delta b$ and $\delta$, which are chosen randomly according to the above distributions. We then numerically minimized $G_A$ over the driving parameters, $\Omega_k$ and $\delta_k$.  The numerical minimization resulted in the following driving parameters (all in units of kHz): $\Omega_1^\star=2\pi\times 225.3$, $\delta_1^\star=2\pi\times 95.6$, $\Omega_2^\star=2\pi\times 13.6$, $\delta_2^\star=2\pi\times 5$, $\Omega_3^\star=2\pi\times 93.6$, $\delta_3^\star=2\pi\times 27.2$, and $\Omega_4^\star=2\pi\times 14.8$, $\delta_4^\star=2\pi\times 25.6$.
\subsection{Fixing the Driving parameters}
When fixing the driving parameters we must take into account the effect of the $S_{1/2}$ ($D_{5/2}$) drive on the $D_{5/2}$ ($S_{1/2}$) states (Section \ref{ME_sec}), as well as the effect of the counter-rotating terms of a drive (Section \ref{BS_sec}) simultaneously. We start by parametrically solving for a general detuned drive the Bloch-Siegert correction equation, $2 \omega_i = \frac{1}{2}\sqrt{\left(\omega_i+\omega_0\right)^2+\Omega_i^2} - x_i \tilde{\Omega}_i$, where $x_i$ would correspond to the ratio between an optimal drive detuning and Rabi frequency, that is $x_i=\frac{\delta_i^\star}{\Omega_i^\star}$. We denote the parametric solution of $\omega_i$ by $\omega_{sol}$. Then, in order to fix the driving parameters of the first driving fields,  $\omega_1$,$\Omega_1$,$\omega_3$, and $\Omega_3$, we numerically solve the following four equations:
\begin{eqnarray}
\omega_1 &=& \omega_{sol}\left(E_D\left(\omega_3,\Omega_3\right),\Omega_1^\star,\delta_1^\star,\Omega_1\right),\\
\omega_3 &=& \omega_{sol}\left(E_S\left(\omega_1,\Omega_1\right),\Omega_3^\star,\delta_3^\star,\Omega_3\right),\\
\Omega_1^\star &=&  \tilde{\Omega}_1\left(\omega_1,E_D\left(\omega_3,\Omega_3\right),\Omega_1^\star,\delta_1^\star,\Omega_1\right),\\
\Omega_3^\star &=&  \tilde{\Omega}_3\left(\omega_3,E_S\left(\omega_1,\Omega_1\right),\Omega_2^\star,\delta_3^\star,\Omega_3\right).
\end{eqnarray} 
For the case of $\mu_B B = 5$~MHz (so  $g_s\mu_B B = 10$~MHz), this results in
$\omega_1=5.904881$~MHz, $\omega_3=9.980794$~MHz, $\Omega_1=225.3107$~kHz, and $\Omega_3 = 93.6311$~kHz. The ideal driving parameters, without the above corrections are given by $\omega_1=5.904411$~MHz, $\omega_3=9.972789$~MHz, $\Omega_1=225.3035$~kHz, and $\Omega_3 = 93.6306$~kHz.
For the second driving fields we neglect the cross-driving effect (modification of the energy gap) and take into account only the  Bloch-Siegert corrections. We obtain that $\omega_2 = 160.5892$~kHz, $\omega_4=72.0503$~kHz, $\Omega_2 = 13.6373$~kHz, and $\Omega_4 = 14.8157$~kHz.

\subsection{Simulations}
We simulated the full driving Hamiltonian of the system with
\begin{eqnarray}
H &=& g_s \mu_B B s_z + g_s \mu_B B\delta b s_z +  g_s \Omega_3\left(1+\delta\right) \cos\left[\omega_3 t\right] s_x + g_s \Omega_4\left(1+\delta\right) \cos\left[\left(\omega_3 t\right)+\frac{\pi}{2}\right]\cos\left[\omega_4 t\right] s_x	 \nonumber \\
  &+& g_s \Omega_1\left(1+\delta\right) \cos\left[\omega_1 t\right] s_x + g_s \Omega_2\left(1+\delta\right) \cos\left[\left(\omega_1 t\right)+\frac{\pi}{2}\right]\cos\left[\omega_2 t\right] s_x \nonumber \\
  &+& g_d \mu_B B S_z + g_d \mu_B B\delta b S_z +  g_d \Omega_1\left(1+\delta\right) \cos\left[\omega_1 t\right] S_x + g_d \Omega_2\left(1+\delta\right) \cos\left[\left(\omega_1 t\right)+\frac{\pi}{2}\right]\cos\left[\omega_2 t\right] S_x	 \nonumber \\
  &+& g_d \Omega_3\left(1+\delta\right) \cos\left[\omega_3 t\right] S_x + g_d \Omega_4\left(1+\delta\right) \cos\left[\left(\omega_3 t\right)+\frac{\pi}{2}\right]\cos\left[\omega_4 t\right] S_x \nonumber\\
  &+& \left[3 S_z^2-\mathrm{I}\frac{5}{2}\left(\frac{5}{2}+1\right)\right] \left(Q + Q_T\right),
\end{eqnarray} 
where $Q$ and $Q_T$ represent the quadrupole and tensor shifts respectively. 
First, we simulated the system with $\delta b=0$, $\delta=0$, and $Q = Q_T = 0$ for a time duration of $0.5$ sec, where we initialized the system in the equal superposition of $\frac{1}{\sqrt{2}}\left(\ket{S_1}+\ket{D_1}\right)$, where $\ket{S_1}$ and $\ket{D_1}$ are the double dressed $S_{1/2}$ and $D_{5/2}$ states with the smallest positive eigenvalue respectively. This gave us a reference for the non-shifted transition frequency, $\nu_{S-D}$ . We then simulated $4927$ realizations with $\delta b \sim \mathcal{N}\left(0,0.5\right)$~kHz, $\delta \sim \mathcal{N}\left(0,4 \times 10^{-4}\right)$, $Q \sim \mathcal{N}\left(-1,1\right)$~Hz, and $Q_T \sim \mathcal{U}\left(-3,0\right)$~Hz. Each simulation resulted in a frequency shift $\Delta \nu$ with respect to $\nu_{S-D}$. The histogram of $\Delta \nu$ is shown in Fig. (\ref{sim}) in the main text. 

\bibliography{refs}

\end{document}